\documentclass[aps,pra,reprint,amsmath,amssymb,
longbibliography,nofootinbib,floatfix,superscriptaddress]{revtex4-1}

\usepackage{graphicx}
\usepackage{physics}
\usepackage{xcolor}
\usepackage{comment}
\usepackage{makecell}
\usepackage{hyperref}
\usepackage{bm}
\usepackage{letltxmacro}

\LetLtxMacro{\originaleqref}{\eqref}

\newcommand{\red}[1]{{\color{black}{#1}}}

\begin{document}

\title{
Wave Vortices Around Oscillating Subwavelength Holes: Water-Wave Observation
}

\author{Junyi Ye}
\affiliation{State Key Laboratory of Surface Physics, Key Laboratory of Micro- and Nano-Photonic Structures (Ministry of Education), and Department of Physics, Fudan University, Yangpu District, Shanghai, 200433, China}

\author{Zheyi Li}
\affiliation{State Key Laboratory of Surface Physics, Key Laboratory of Micro- and Nano-Photonic Structures (Ministry of Education), and Department of Physics, Fudan University, Yangpu District, Shanghai, 200433, China}

\author{Alexey Y. Nikitin}
\affiliation{Donostia International Physics Center (DIPC), Donostia-San Sebasti\'an 20018, Spain}
\affiliation{IKERBASQUE, Basque Foundation for Science, Bilbao 48009, Spain}

\author{Franco Nori}
\affiliation{RIKEN Center for Quantum Computing, RIKEN, Wako-shi, Saitama 351-0198, Japan}
\affiliation{Physics Department, University of Michigan, Ann Arbor, Michigan 48109-1040, USA}

\author{Wenzhe Liu}
\email{wzliu@fudan.edu.cn}
\affiliation{State Key Laboratory of Surface Physics, Key Laboratory of Micro- and Nano-Photonic Structures (Ministry of Education), and Department of Physics, Fudan University, Yangpu District, Shanghai, 200433, China}

\author{Konstantin Y. Bliokh}
\email{konstantin.bliokh@dipc.org}
\affiliation{Donostia International Physics Center (DIPC), Donostia-San Sebasti\'an 20018, Spain}
\affiliation{IKERBASQUE, Basque Foundation for Science, Bilbao 48009, Spain}

\author{Lei Shi}
\email{lshi@fudan.edu.cn}
\affiliation{State Key Laboratory of Surface Physics, Key Laboratory of Micro- and Nano-Photonic Structures (Ministry of Education), and Department of Physics, Fudan University, Yangpu District, Shanghai, 200433, China}


\begin{abstract}
We consider a two-dimensional wave system containing a subwavelength hole, such as an aperture in an interface supporting surface electromagnetic or acoustic waves, or an island in a fluid surface sustaining gravity-capillary waves. 
Recent studies have revealed the emergence of pronounced wave vortices around such structures, termed type-II vortices, in contrast to conventional (type-I) vortices associated with phase singularities and intensity nulls.
A striking natural manifestation of type-II vortices occurs in ocean tides around islands such as New Zealand, Madagascar, and Iceland, where the tidal phase increases by $\pm 2\pi$ around the island.
Although this phenomenon is usually associated with the Coriolis effect from the rotation of the Earth, here we demonstrate the controlled generation of type-II vortices using a minimal and tunable setup: a dipole-oscillating subwavelength hole and a single incident plane wave.
Using laboratory gravity-capillary waves and an oscillating subwavelength `island', we directly measure the resulting phase structure, topological charge, and wave angular momentum. We show that the emergence and handedness of the vortices can be precisely controlled via the relative phase between the dipolar source and the incident wave.
Our results offer a simple and versatile mechanism for engineering subwavelength wave vortices, with potential applications in a variety of two-dimensional wave systems.
\end{abstract}

\maketitle


\section{Introduction}

Wave vortices are generic topological objects in complex wave fields, characterized by phase winding and orbital angular momentum (OAM) \cite{Nye1974, Bazhenov1990, Ceperley1992AJP, Allen1992PRA}. In two-dimensional (2D) wave systems, vortices typically appear around {\it phase singularities}, i.e., points where the wave intensity vanishes and phase becomes undefined \cite{Nye1974}. The phase accumulates by $2\pi\ell$ when encircling such points, where $\ell$ is an integer {\it topological charge} of the vortex. Vortex waves of this type have been extensively studied and exploited across a wide range of physical platforms, including optics \cite{Allen_book, Torres_book, Dennis2009PO}, acoustics \cite{Hefner1999JASA, Guo2022JAP}, plasmonics \cite{Ohno2006OE, Prinz2023ACSP}, quantum electron waves \cite{Bliokh2017PR}, ocean tidal waves \cite{Nye1988, Berry2001}, and water waves \cite{Smirnova2024PRL, Wang2025N}.

In homogeneous media, vortices with phase singularities naturally arise from the interference of regular or random wave \cite{Masajada2001OC, BerryDennis2000}. A distinct class of vortices emerges in the presence of defects in the 2D plane. Indeed, when the wavefield is defined only outside a finite excluded area (in what follows, a {\it hole} in the plane), the phase can wind around this hole {\it without phase singularities}. We term these structures ``type-II vortices'', in contrast to conventional ``type-I vortices'' associated with phase singularities \cite{Domina2025}. 

A prominent example of the two types of vortices is provided by ocean tidal waves. Phase singularities in ocean tides, accompanied by type-I vortices with $\ell=\pm 1$, are known as amphidromic points \cite{Cartwright_book, Nye1988, Berry2001}. In turn, islands represent natural holes in the ocean, around which the tidal phase can wind by $\pm 2\pi$, as observed near New Zealand, Madagascar, and Iceland \cite{Bye1975, Heath1977, Heath1985, Walters2001, Stevens2021, Domina2025}. These are type-II tidal-wave vortices with $\ell=\pm 1$. In optics and acoustics, similar vortices can appear around apertures or defects in interfaces supporting surface waves \cite{Gorodetski2010PRB, Vanacore2019NM, Triolo2019SR, Domina2025}.

A characteristic feature of type-II vortices is their ability to concentrate wave energy and OAM on a {\it subwavelength} scale around the hole (note that the islands of New Zealand, Madagascar, and Iceland are much smaller than the typical tidal wavelength $\lambda \sim \!10,\!000\,$km). Thus, such vortices are formed by {\it near-fields} around the hole and cannot be described by the interference of propagating plane waves. 

In this work, we demonstrate the first controllable generation of type-II vortices in laboratory water waves via interference between a dipole-oscillating subwavelength `island' and a single incident plane wave. In contrast to tidal vortices around oceanic islands, whose origin is commonly attributed to the Coriolis effect \cite{Longuet-Higgins1967, Longuet-Higgins1969}, our system possesses no intrinsic chirality. Instead, the vortex emerges from the interference of two linear wave components and is fully controlled by their relative phase and amplitude. 
We theoretically analyze and experimentally measure the topological and OAM properties of the generated type-II vortices. 

Owing to its conceptual simplicity, the proposed mechanism is readily extendable to a wide range of 2D wave systems containing subwavelength holes. Our results provide an accessible platform for studying type-II vortices, and may also offer new insight into tidal-wave dynamics around islands, where horizontal dipole-like oscillations of the island can be induced by Earth tides \cite{Melchior1974} coherent with ocean tides.


\section{Theoretical model}

We consider linear monochromatic scalar waves in a 2D plane containing a circular hole of radius $a$: ${\bf r}\equiv(x,y) \in \mathbb{R}^2\backslash D_a$, where $D_a =\{r=\sqrt{x^2+y^2}<a \}$. The real wavefield can be written as $\Psi({\bf r},t) = {\rm Re}\!\left[\psi({\bf r})e^{-i\omega t}\right]$, where $\psi({\bf r})$ is the complex wavefield with intensity $|\psi|^2$ and phase ${\rm arg}(\psi)$, and $\omega$ is the wave frequency. We assume that the wavefield satisfies the Helmholtz equation $\nabla^2 \psi+k^2 \psi =0$, where the wavenumber $k$ is related to the frequency via the dispersion relation $\omega = \omega(k)$.

\begin{figure}[t]
\centering
\includegraphics[width=\linewidth]{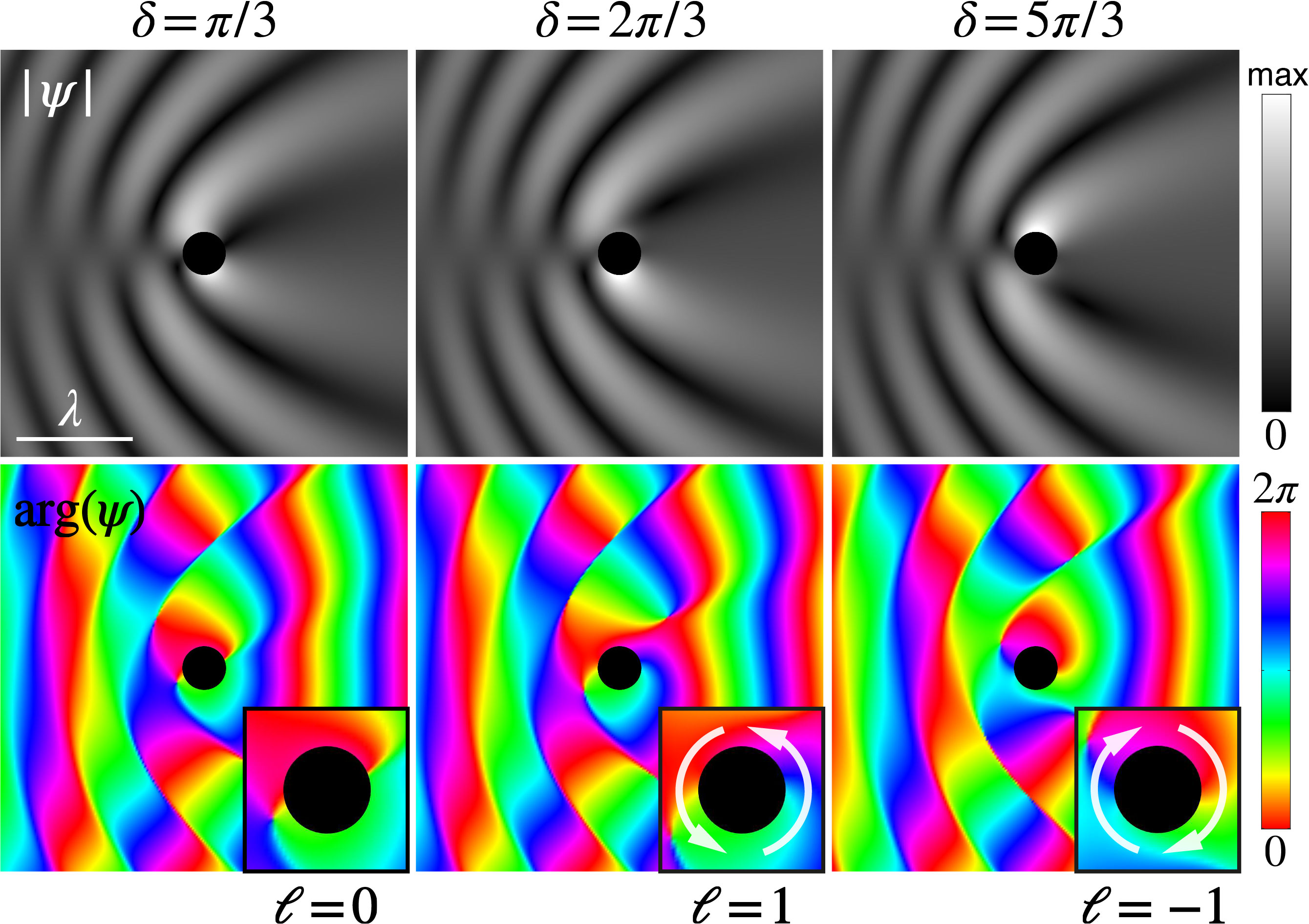}
\caption{Amplitude and phase distributions of the complex wave field \eqref{eq:interference} for $A=0.3$, $ka=1.2$, and different values of the phase $\delta$. Type-II vortices with topological charges $\ell=1$ and $\ell=-1$ can be seen in the phase patterns around the hole for $\delta = 2\pi/3$ and $\delta = 5\pi/3$, respectively. 
}
\label{Fig_1}
\end{figure}

The interference between a linear-dipole field emitted by the hole along the $\pm y$ directions and an $x$-propagating plane wave can be written as
\begin{equation}
\label{eq:interference}
\psi ({\bf r}) \propto A e^{ikx+i\delta} + \frac{y}{r}H_1^{(1)}\!(kr)\,,
\end{equation}
where $A$ and $\delta$ denote the amplitude and phase of the plane wave relative to the dipole, and $H_1^{(1)}$ is the Hankel function of the first kind. (A simplified version of this model with $H_1^{(1)}\!(kr) \simeq -2i/(\pi kr)$, $kr \ll 1$, was considered in Ref.~\cite{Domina2025}; here we employ the exact expression, which considerably affects the results below.) The field \eqref{eq:interference}, shown in Fig.~\ref{Fig_1}, satisfies the Helmholtz equation, and its dipole part contains only outgoing waves because $H_1^{(1)}\!(kr) \sim (kr)^{-1/2}e^{ikr}$ for $kr \gg 1$. 
We assume a subwavelength hole size, $a \lesssim 1/k = \lambda/2\pi$, and not large plane-wave amplitude $A \lesssim 1$. 
In this regime, the dipole field dominates at the hole boundary (assuming that it satisfies appropriate boundary conditions), allowing us to neglect the wave scattered by the hole from the incident plane wave (see the Supplementary Materials). 

\begin{figure}[t]
\centering
\includegraphics[width=\linewidth]{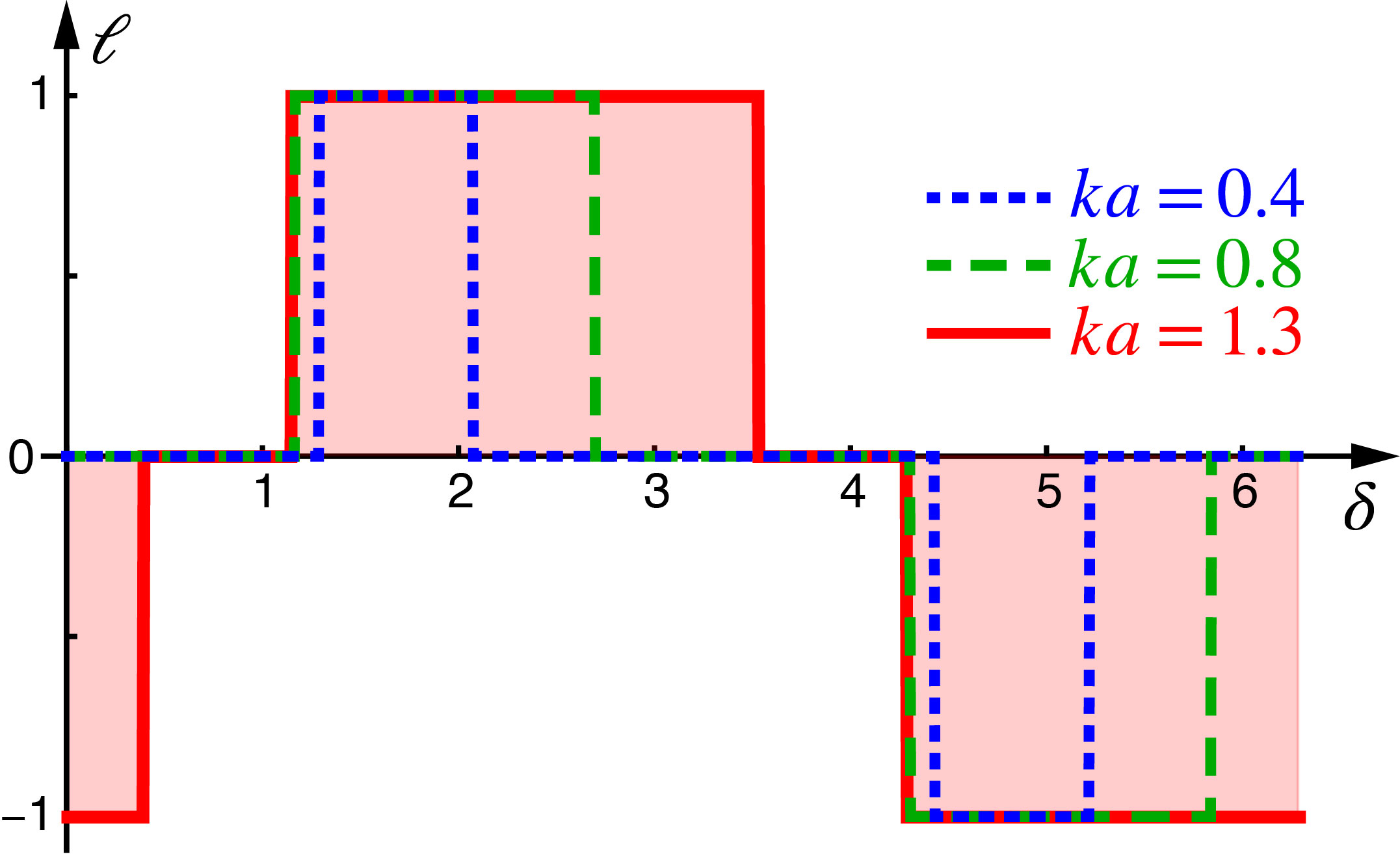}
\caption{Topological charge \eqref{eq:topological} as a function of the phase $\delta$ calculated around the holes of different radii $a$ for the field \eqref{eq:interference} with $A=0.3$. The red curve corresponds approximately to the parameters used in Fig.~\ref{Fig_1}.}
\label{Fig_2}
\end{figure}

Figure~\ref{Fig_1} presents the amplitude and phase distributions of the field \eqref{eq:interference} for $A=0.3$ and different values of the phase $\delta$. Since expression \eqref{eq:interference} does not explicitly depend on the hole radius $a$, the hole can be introduced by excluding the corresponding region $D_a = {r<a}$ from these distributions. The presence of a type-II vortex around the hole is quantified by the topological charge
\begin{equation}
\label{eq:topological}
\ell = \frac{1}{2\pi}\oint_{\partial D_a}\boldsymbol{\nabla}{\rm arg}(\psi) \cdot d{\bf r} \,,
\end{equation}
which measures the phase winding along the hole boundary.
Figure~\ref{Fig_2} shows the dependence of the topological charge \eqref{eq:topological} on the phase $\delta$ for the chosen amplitude $A$ and different hole radii $a$. 
For a certain range of $\delta$, type-II vortices with $\ell=1$ occur, whereas vortices with $\ell=-1$ appear upon shifting the phase by $\pi$.
This symmetry follows from the fact that the transformations $\delta \to \delta + \pi$ and $y \to -y$ in Eq.~\eqref{eq:interference} produce the same field up to an overall $\pi$ phase.

\begin{figure}[t]
\centering
\includegraphics[width=\linewidth]{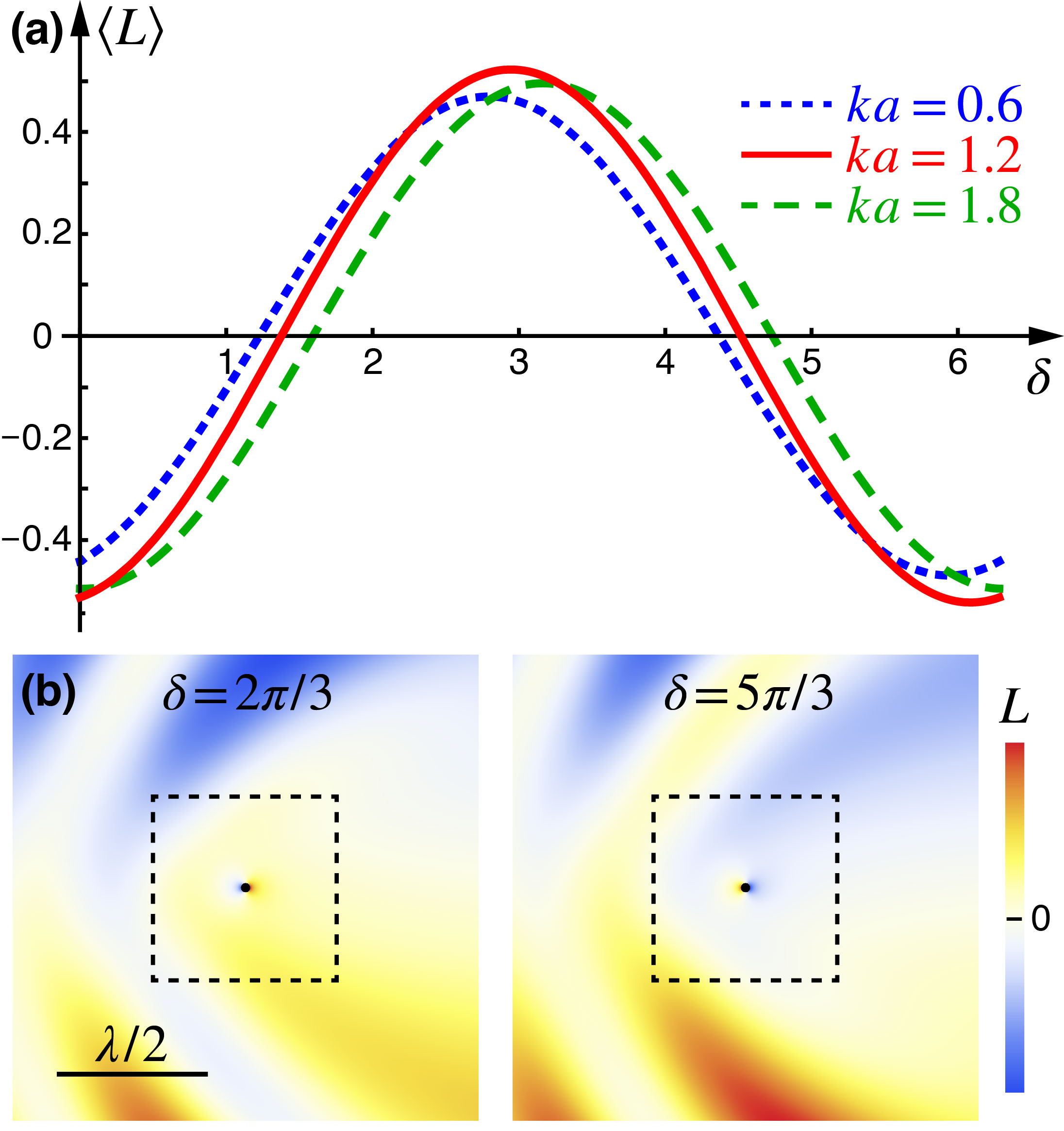}
\caption{(a) Normalized OAM \eqref{eq:OAM} versus phase $\delta$ for the field \eqref{eq:interference} with $A=0.3$ and different hole sizes $a$. The integration area is a symmetric square excluding the hole: $S= \{|x| < 2.86/k, |y|<2.86/k\} \backslash D_a$. The dependence $\langle L\rangle (\delta)$ correlates with the topological charge $\ell (\delta)$ in Fig.~\ref{Fig_2}. (b) Spatial distributions of the OAM density $L({\bf r})$ for $\delta = 2\pi/3$ and $5\pi/3$, corresponding to $\ell=1$ and $\ell=-1$ vortices in Fig.~\ref{Fig_1}. The dashed squares indicate the OAM integration area $S$. The hole size is reduced to $ka = 0.1$ to expose the deep-subwavelength structure near the origin.}
\label{Fig_3}
\end{figure}

While Eq.~\eqref{eq:topological} provides the topological index of the vortex around the hole, its dynamical properties can be characterized by the OAM of the field. Since the field \eqref{eq:interference} is not square-integrable over the entire plane (implying infinite energy and OAM), we evaluate a normalized OAM within a finite symmetric area $S$ surrounding the hole \cite{Allen_book}:    
\begin{equation}
\label{eq:OAM}
\langle L \rangle = \frac{\int_S \psi^* \hat{L} \psi\, d^2{\bf r}}{\int_S \psi^* \psi\, d^2{\bf r}}\,,
\end{equation}
where $\hat{L} = -i(x\,\partial/\partial y - y\,\partial/\partial x) = -i \partial/\partial\varphi$ is the OAM operator, with $\varphi$ being the azimuthal coordinate in the ${\bf r}$ plane. 
Figure~\ref{Fig_3}(a) displays the dependence of the OAM \eqref{eq:OAM} on $\delta$ for fixed amplitude $A$ and different hole sizes $a$. 
Analytical calculation using Eqs.~\eqref{eq:interference} and \eqref{eq:OAM} shows that the OAM has the form $\langle L \rangle(\delta) \propto -\cos(\delta + \gamma)$, where the phase offset $\gamma$ depends on the choice of the integration area $S$ (see the Supplementary Materials). 
(For the plots in Fig.~\ref{Fig_3}, $|\gamma| \ll 1$, whereas the simplified model of Ref.~\cite{Domina2025} yields $\gamma = \pi/2$.) 

Comparison of Figs.~\ref{Fig_2} and \ref{Fig_3}(a) reveals a clear correlation between the OAM and the topological charge \eqref{eq:topological}, confirming that the OAM serves as a dynamical signature of type-II vortices. In addition, Fig.~\ref{Fig_3}(b) shows the spatial distribution of the OAM density
$L({\bf r}) = {\rm Re}\!\left[\psi^*({\bf r})\, \hat{L}\, \psi({\bf r}) \right]$ around the hole for the $\ell=\pm 1$ vortices.


\section{Water-wave experiment}

\begin{figure}[t]
\centering
\includegraphics[width=\linewidth]{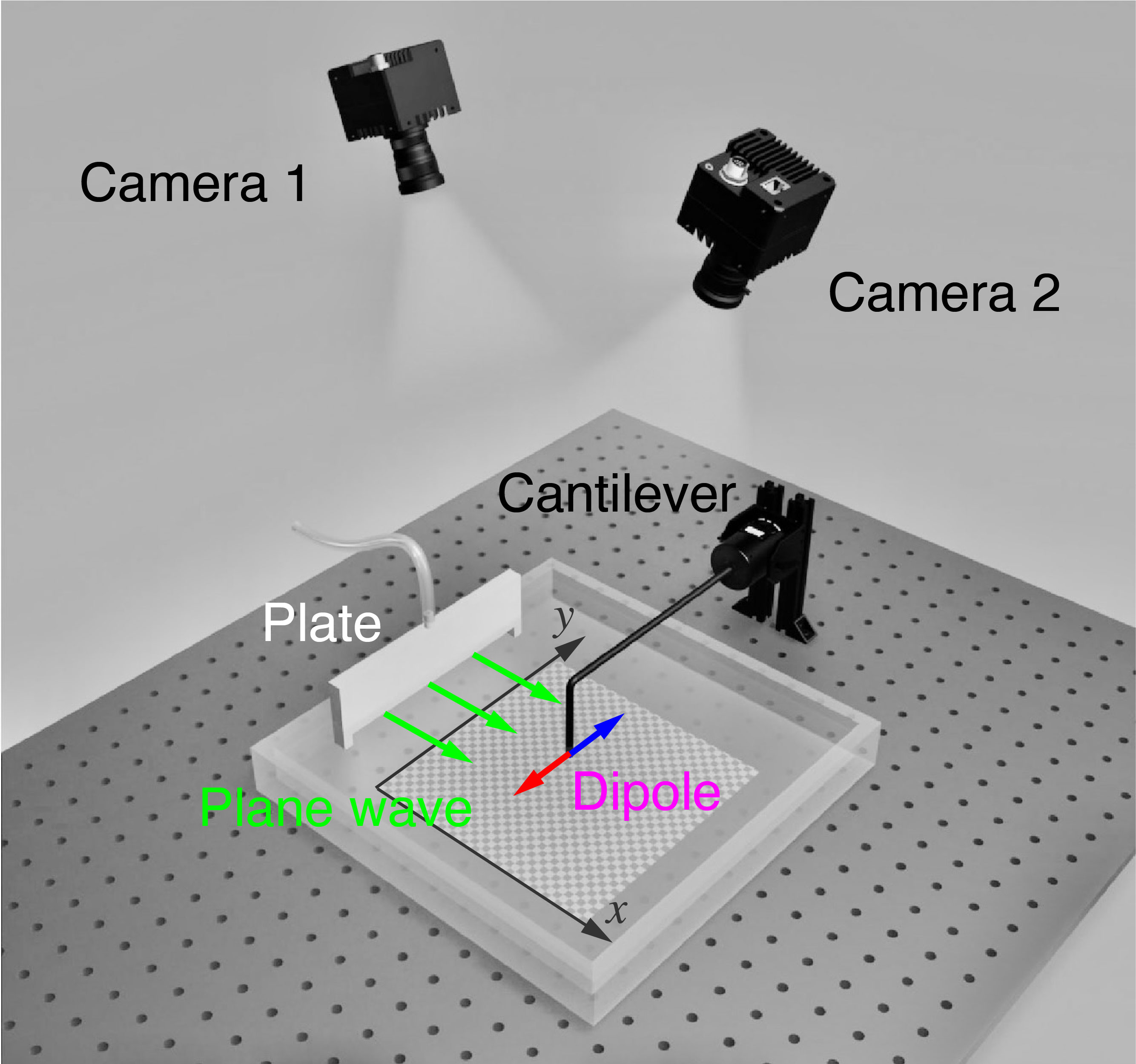}
\caption{Schematic of the experimental setup (see explanation in the text).}
\label{Fig_4}
\end{figure}

The experiment was performed using the setup shown in Fig.~\ref{Fig_4} and involving a $40 \times 40 \text{ cm}^2$ wave tank with depth $h=1.5 \text{ cm}$. The driving frequency was set to $\omega/2\pi = 6.2 \text{ Hz}$, corresponding to a gravity-capillary wavelength of $\lambda \simeq 4.4 \text{ cm}$.
Two excitation mechanisms were employed: a speaker coupled to a 3D-printed hollow plate via a flexible hose to generate a plane wave, and an L-shaped hollow steel cantilever (diameter $d = 1.2\;$cm) driven by a modal shaker to produce a dipolar excitation. When immersed, the cantilever acts as a subwavelength “island,” effectively defining the hole in the water surface. The distance between the plane-wave source and the cantilever center was fixed at $3\lambda$.
\red{However, an initial phase offset still exists due to the different excitation devices used for the two sources. Therefore, we introduce an initial phase to compensate for this phase difference
(see the Supplementary Materials).}


To overcome the occlusion introduced by the cantilever along the $y$ direction, a dual-camera imaging system was used. First, the recorded images were transformed into a top-down (orthographic) view via a perspective-correction algorithm, calibrated using a $30 \times 30,\text{cm}^2$ reference grid placed at the bottom of the tank \red{(see Supplementary Materials)}. Second, the surface elevation fields $\Psi({\bf r},t)$ from the two cameras 
were reconstructed using the Fast Checkerboard Demodulation (FCD) method \cite{Wildeman2018, Wang2025N}.

Third, temporal filtering was applied to the reconstructed fields to suppress stray waves and higher-harmonic components, which are unavoidable in water waves \red{(see Supplementary Materials)}. Specifically, a Fast Fourier Transform (FFT) was performed on the temporal signal at each spatial point ${\bf r}$, followed by the application of a Gaussian frequency window centered at 6.2~Hz with a $1/e$ half-width of 0.2~Hz. 
\red{To achieve sufficient frequency resolution in the
FFT domain, the temporal evolution of the field $\Psi ({\bf r},t)$ was recorded for 5 seconds.}

Fourth, to ensure accurate alignment between the two camera views, a cross-correlation procedure was used to identify a common reference point \red{(see Supplementary Materials)}. Around this point, $16 \times 16\,\text{cm}^2$ sub-regions were extracted from each image and subsequently stitched together, thereby eliminating the visual gap caused by the cantilever along the $y$ axis.
Finally, the complex field $\psi({\bf r})$ was obtained via a Hilbert transform of the real field $\Psi({\bf r},t)$.

\begin{figure}[t]
\centering
\includegraphics[width=\linewidth]{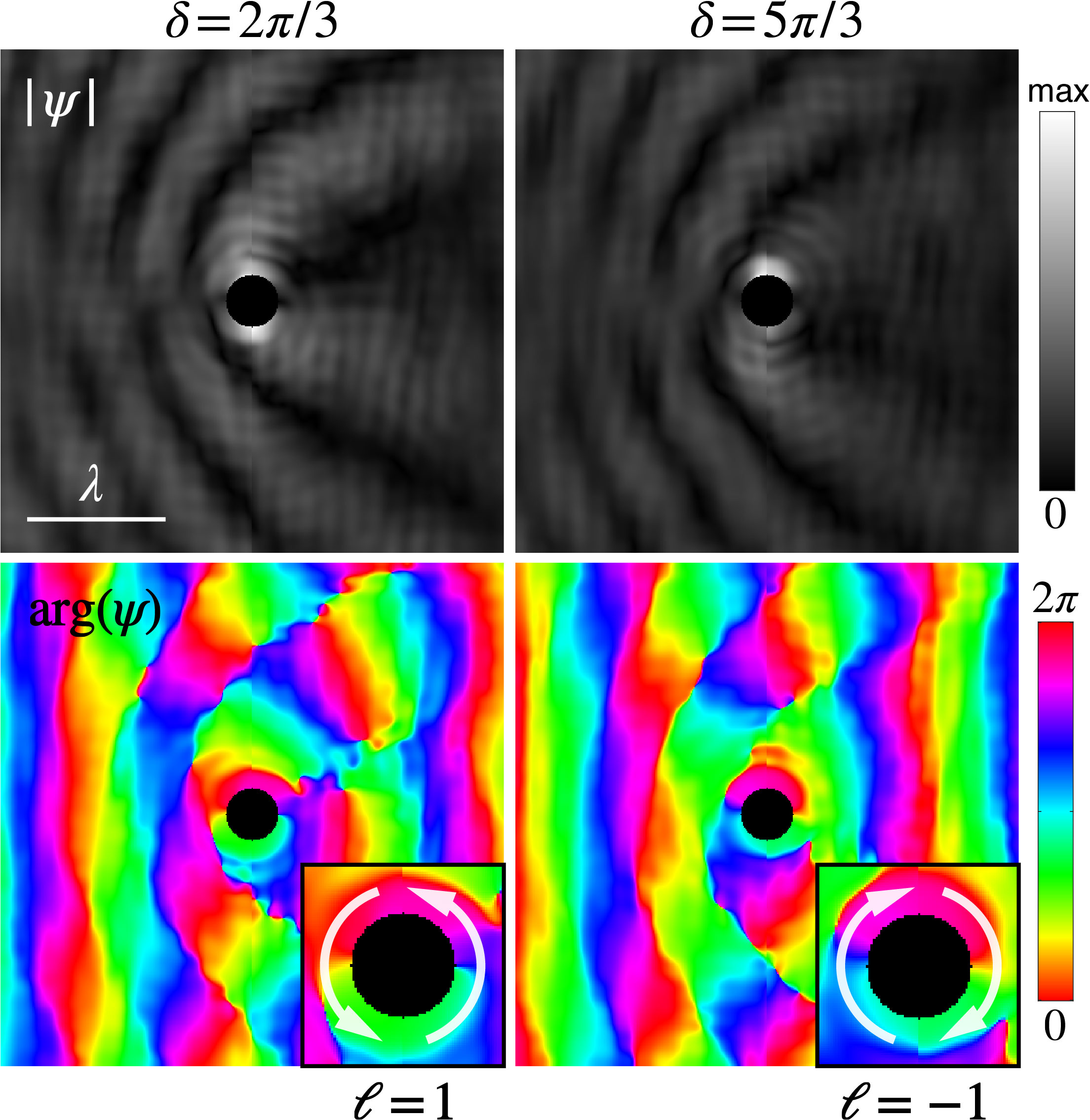}
\caption{Experimentally measured amplitude and phase distributions of the water-wave field $\psi({\bf r})$, exhibiting type-II vortices with $\ell=1$ and $-1$ for $\delta=2\pi/3$ and $5\pi/3$, respectively. The parameters approximately match those in Fig.~\ref{Fig_1}: $A \simeq 0.3$, $ka \simeq 1.2$.}
\label{Fig_5}
\end{figure}

Figure~\ref{Fig_5} presents the experimentally measured amplitude and phase distributions of the field $\psi({\bf r})$ for parameters corresponding to the theoretical results in Fig.~\ref{Fig_1}. The emergence of type-II vortices with $\ell = \pm 1$ is clearly observed, together with field patterns which are excellent agreement with the theoretical model~\eqref{eq:interference}. \red{Furthermore, by analyzing the spatial Fourier spectrum of the measured field and applying a cut-off filter for wavenumbers higher than $1.4k$, we confirm that the high-intensity vortex field around the hole is essentially formed by high-wavenumber near-field components (see Supplementary Materials).}

Figures~\ref{Fig_6}(a) and (b) show the dependencies of the topological charge \eqref{eq:topological} and OAM \eqref{eq:OAM}, calculated from the experimental data, on the phase $\delta$. 
\red{(Each $\langle L \rangle$ or $\ell$ value for a given $\delta$ was obtained by averaging 50 neighboring frames sampled at a frequency of 100~Hz, which corresponds to approximately three oscillation cycles.)}
These plots closely match the theoretical predictions in Figs.~\ref{Fig_2} and \ref{Fig_4}(a) (up to a slight vertical offset in the OAM curve).   

Thus, experimental observations fully confirm the appearance and key properties of type-II vortices around subwavelength dipole-oscillating holes, as described by the theoretical model \eqref{eq:interference}. 

\begin{figure}[t!]
\centering
\includegraphics[width=\linewidth]{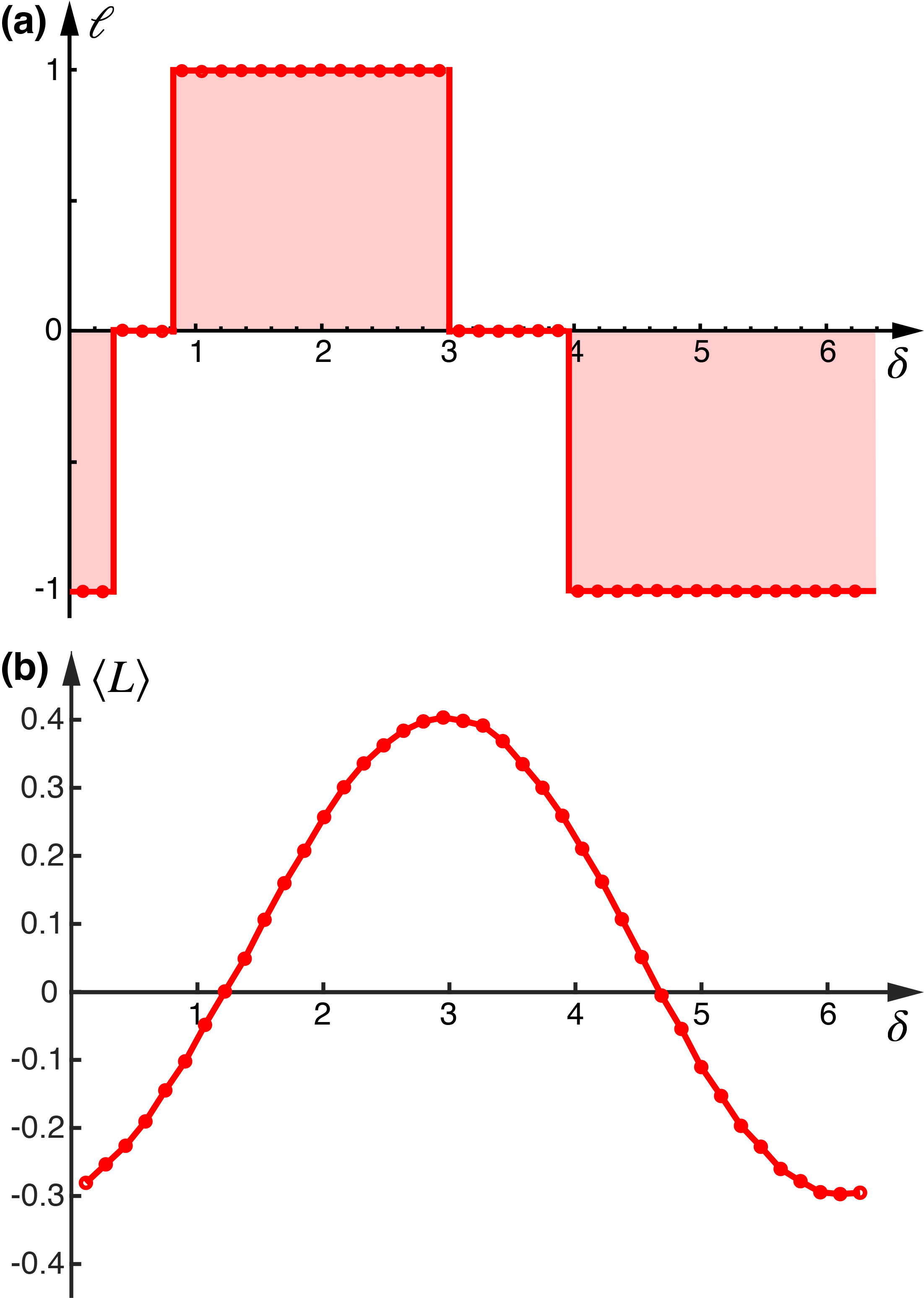}
\caption{(a) Topological charge \eqref{eq:topological} as a function of phase $\delta$, calculated from the experimentally measured water-wave field $\psi({\bf r},t)$. The parameters correspond approximately to the red curve in Fig.~\ref{Fig_2}: $A\simeq 0.3$ and $ka \simeq 1.3$. 
(b) Normalized OAM \eqref{eq:OAM} versus $\delta$ calculated for the experimentally measured water-wave field. The parameters correspond approximately to the red curve in Fig.~\ref{Fig_3}: $A\simeq 0.3$, $ka \simeq 1.2$, and the integration area $S= \{|x| < 2.86/k, |y|<2.86/k\} \backslash D_a$.}
\label{Fig_6}
\end{figure}





\section{Conclusions}

In summary, we have demonstrated the controlled generation of type-II wave vortices around subwavelength holes in a two-dimensional water-wave system. Using a minimal interference scheme between a dipole-oscillating ``island'' and a single incident plane wave, we showed that vortex states with topological charges $\ell = \pm 1$ can be created and tuned via the relative phase and amplitude of the two wave components.
Our results reveal that robust vortex behavior can emerge without phase singularities, originating instead from phase winding and near-fields around subwavelength holes. 
The topological and dynamical properties of such vortices are quantified by the phase-winding number and OAM of the field, whereas excellent agreement between theory and water-wave experiments validates the minimal interference model.


Beyond the specific water-wave implementation, the simplicity and generality of the proposed mechanism make it readily applicable to a wide range of two-dimensional wave systems, including optical, acoustic, and plasmonic platforms supporting surface waves. In this context, the ability to engineer subwavelength vortices without relying on intrinsic material chirality or complex structured fields provides a versatile tool for controlling wave energy flow and angular momentum at small scales.
Finally, our results offer a new perspective on naturally arising type-II vortices around islands, by highlighting the role of near-fields and effective dipolar responses. 

\vspace{1cm}

\begin{acknowledgments}
L. S. is supported by National Key R\&D Program of China (2023YFA1406900); National Natural Science Foundation of China (No. 12321161645, No. 12234007, No. 12221004, No. T2394480, T2394481); National Key R\&D Program of China (2022YFA1404800); Science and Technology Commission of Shanghai Municipality (2019SHZDZX01 and 23DZ2260100).
K. Y. B. is supported by the Marie Sk\l{}odowska-Curie COFUND Programme of the European Commission (project HORIZON-MSCA-2022-COFUND-101126600-SmartBRAIN3). 
A. Y. N. is supported by the Department of Science, Universities
and Innovation of the Basque Government (grant PIBA-2023-
1-0007 and the IKUR Strategy) and the Spanish Ministry of
Science and Innovation (grant PID2023-147676NB-I00).
F. N. was supported in part by the Japan Science and Technology Agency (JST) [via the CREST Quantum Frontiers program Grant No. JPMJCR24I2, and the ASPIRE program (Grant Number JPMJAP2513)].
 
\end{acknowledgments}

\bibliography{refs}

\end{document}